\documentclass[11pt,twoside]{article}
\usepackage{asp2010}

\resetcounters

\begin{document}

\title{Another Unsung Lowell Observatory Achievement: The First Infrared Observation of a Comet}
\author{J.~N. Marcus
\affil{Friends of Lowell Observatory, 19 Arbor Road, St. Louis, MO, 63132, USA}}

\begin{abstract}
Carl Lampland was the first to observe a comet in the infrared, a feat little
known today because he failed to formally publish his data.  I have retrieved
the radiometry of this comet, C/1927 X1 (Skjellerup-Maristany), taken in broad
daylight, from Lampland's logbook in the Lowell Observatory archives, and
present a preliminary reduction of it here.  There are similarities between
Lampland's pioneering achievement and V.~M. Slipher's discovery of the redshifts
of the spiral nebulae (and thus, arguably, the expansion of the Universe).  Each
astronomer used state-of-the-art instrumentation, received rave reviews at
American Astronomical Society meetings where their novel data were presented,
and suffered under-recognition in ensuing decades.  A common thread in these
poor outcomes was their lackadaisical approach to formal publication -- in
Slipher's case, publishing in internal or secondary outlets, and in Lampland's
case, not publishing at all.  \end{abstract}

\section{Introduction}

The first infrared observation of a comet is widely and wrongly attributed
\citep{2000Crovisier,1983Encrenaz,1993Festou,1997Gehrz,1991Hanner,1981Hobbs,1982Ney,2001Sekanina}
to that of C/1965 S1 (Ikeya-Seki) by 
\citet{1966Becklin}.
But as 
\citet{1984Hoag}
and 
\citet{1991Yeomans}
have pointed out, it is Lowell Observatory staff astronomer Carl O.\ Lampland
(1873-1951) (Figure 1) who must claim that honor.  The achievement fell into
obscurity because Lampland never converted his abstract report
\citep{1928Lampland}
into a formal paper.  I have retrieved and examined the original radiometry data
on the comet, C/1927 X1 (Skjellerup-Maristany; old style 1927k = 1927 IX), from
Lampland's logbook in the Lowell Observatory archives.  Here I sketch a method
to reduce these data, and present preliminary results to demonstrate that the
data are indeed usable.  I then draw parallels between Lampland's pioneering
observations, and the discovery of the redshifts of the spiral nebulae by
V.~M. \citet{1917Slipher}, including the under-recognition of both
astronomers in the astronomical community for their respective achievements.

\begin{figure}[t]
\plotfiddle{'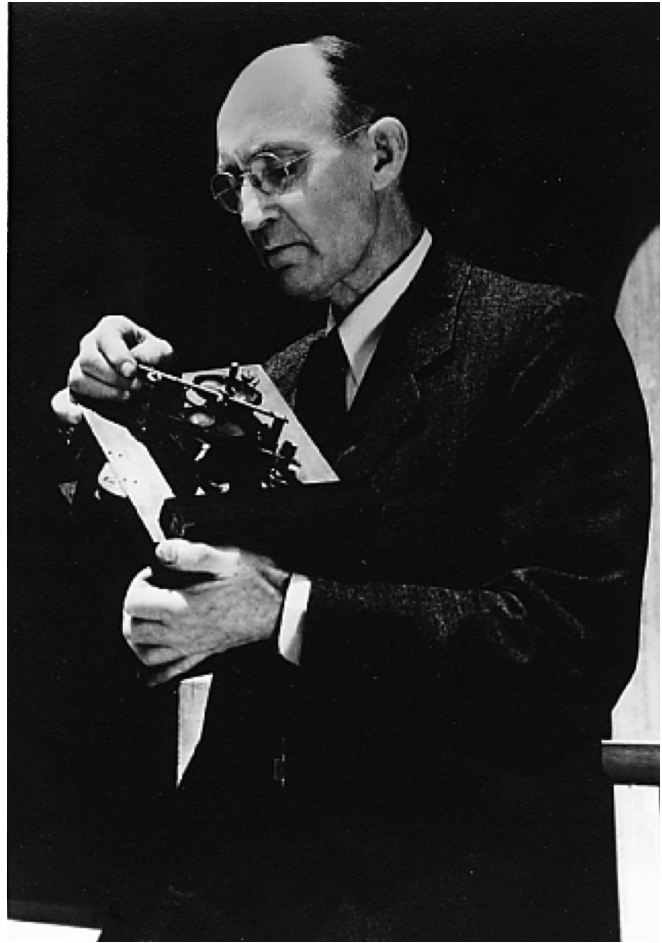'}{8cm}{0}{77}{77}{-120}{-115}
\vskip 1.5truein
\caption{
Carl Otto Lampland (1873-1951) holds a stellar radiometer used in determining the temperatures 
of planets at Lowell Observatory.  He was one of the permanent ``troika'' of staff astronomers 
(with E.~C. and V.~M. Slipher) remaining at Lowell Observatory in the decades after the 
death of Percival Lowell.  All three observed comet C/1927 X1.  Courtesy Lowell Observatory Archives.
}
\end{figure}

\section{The Comet and the Observations}

Several weeks after its discovery, the comet unexpectedly burst into daylight
visibility on Dec.\ 15, taking Lowell Observatory astronomers and the rest of
the world by surprise.  It reached perihelion on 1927 Dec.\ 18.18 Universal Time
(UT) at q = 0.176 astronomical units (AU), having passed nearly between the
earth and the sun on Dec. 15.39 UT at a minimum scattering angle of $\theta =
180^\circ - {\rm phase\ angle} = 6.6^\circ$ (the significance of which will be
discussed).  
Over Dec. 16-19, in broad daylight, E.~C. and V.~M. Slipher took
spectrograms with the 24-inch Clark refractor \citep{1928Slipher},
while \citet{1928Lampland}
obtained radiometry measures with the 42-inch reflector in the neighboring dome, assisted by his wife Verna 
\citep{2003Giclas}.
The observing circumstances are provided in the ephemeris in Table 1.  The
fractional UT dates correspond to the mean times of the observations on each
day.  $\Delta$ and $r$ are the comet's Earth and Sun distances in AU and
$\epsilon$ is the elongation from the sun.  
The radiometric quantities $\varphi_{H_2O}$ and $R_f(\theta)$ will be explained below.

\begin{table}[t]
\caption{Circumstances of Lampland's Radiometric Observations of Comet C/1927 X1}
\smallskip
\begin{center}
{\small
\begin{tabular}{|c|c|c|c|c|c|c|}
\tableline
\noalign{\smallskip}
Mean Date (UT) & $\Delta$ (AU) & $r$ (AU) & $\epsilon$ ($^\circ$) & $\theta$ ($^\circ$) & $\varphi_{H_2O}$ & $R_f(\theta)$ \\
\noalign{\smallskip}
\tableline
\noalign{\smallskip}
Dec. 16.96 & 0.822 & 0.183 & 5.4 & 30.3 & 0.636 & 3.059 \\
Dec. 17.91 & 0.861 & 0.177 & 7.9 & 49.7 & 0.374 & 0.800 \\
Dec. 18.91 & 0.909 & 0.179 & 9.8 & 70.1 & 0.261 & 0.541 \\
Dec. 19.92 & 0.961 & 0.190 & 11.1 & 88.7 & 0.177 & 0.262 \\
\noalign{\smallskip}
\tableline
\end{tabular}
}
\end{center}
\end{table}

\section{The Radiometer}\label{sec:radiometer}

Lampland had been measuring planetary temperatures with the Lowell 42-inch
reflector during the 1920s in collaboration with the pioneering infrared
physicist W.~W. Coblentz (1873-1962), who designed the radiometer
\citep[][Figure 1]{1923Coblentz}.
The full device, which attached to the telescope, contained tiny thermocouples
in an evacuated chamber (see Figures 1 and 2 in \citet{1923Coblentz}).
Different cutoff transmission filters, or ``screens,'' could be interposed in
the light path to parse the radiation and deduce its distribution between light
and heat (see Figures 3 and 4 in \citet{1923Coblentz}).
The screens and their effective cutoff wavelengths -- below which radiation was
transmitted -- were a water cell (1.2 $\mu$m), pyrex glass (2.7 $\mu$m), quartz
(3.8 $\mu$m), microscope coverslip glass (6 $\mu$m), 
and fluorite (12.5 $\mu$m).  The total waveband without any screen was
effectively bounded by atmospheric absorption by ozone below 0.3 $\mu$m, and the
``great wall'' of carbon dioxide above 13.8 $\mu$m, although a trace of infrared
radiation could leak beyond the ``wall'' between 16.8 $\mu$m and the $\sim$18.5
$\mu$m cutoff of the radiometer's rock-salt window.  The amplified
thermoelectric current induced by the comet's radiation was recorded by a
Thompson iron-clad galvanometer.  Lampland and his wife recorded the needle
excursions, to mm precision, with and without the transmission screens, in the
logbook.  These scale directly to flux and are the basic data.

\section{Reduction of Observations}\label{sec:reduction}

Radiation from comet grains is comprised of scattered sunlight,
$f_{scat}(\theta)$, and sunlight absorbed and re-emitted as heat, $f_{emit}$.
$f_{scat}(\theta)$ is a function of the scattering angle, 
$\theta$ (see Section \ref{sec:radiometer}).
Each flux distribution is approximately blackbody in character.  
The effective temperature of $f_{scat}(\theta)$ is $T_{scat} = T_{Sun}C$, where
$T_{Sun} = 5800^\circ$ K is the color temperature of the Sun, and $C$ is a color
index near to or slightly less than 1, depending on the amount 
of reddening of the sunlight by the dust grains.  The effective temperature of
$f_{emit}$ is $T_{emit} = T_{BB}S$, where the blackbody temperature $T_{BB} =
278^\circ r^{-1/2}$, and $S$ is the grain ``superheat.''\footnote{At sizes of the order
of the wavelength of sunlight, $\sim$ 1/2 $\mu$m, coma grains are so small
that they do not radiate away their heat efficiently, so $S \geq 1$; see 
\citet{1982Ney,1997Gehrz}.}
By Wien's law, peak emissions occur at $\lambda = 0.50$ $\mu$m in the visible for $f_{scat}(\theta)$, 
and, at the comet's heliocentric 
distance of $r \sim 0.18$ AU (Table 1), between 3.0 $\mu$m and 4.5 $\mu$m in the near-infrared for 
$f_{emit}$, corresponding to grain temperatures in the range $650^\circ {\rm K} \leq T \leq 1000^\circ {\rm K}$, 
which depend on the grain superheats.  The $f_{scat}(\theta)$ and $f_{emit}$ spectral distributions overlap in 
the near-IR over roughly 1.2 $\mu$m -- 2.3 $\mu$m.  In this circumstance, the water cell measures essentially 
$f_{scat}(\theta)$, while the pyrex and quartz screens (and the unfiltered measurements) capture segments 
of both $f_{scat}(\theta)$ and $f_{emit}$.  Superimposed on the radiation may be atomic or molecular 
emission features, mainly from C$_2$ and Na for $f_{scat}(\theta)$, and silicates for $f_{emit}$.

The basic data element for analysis is the ratio of the fluxes, as measured in the galvanometer, 
with ($\nu^\prime$) and without ($\nu$) a given transmission screen:

\begin{equation}\label{marcuseq1}
\varphi = \frac{\nu^\prime}{\nu}
\end{equation}

\noindent $\varphi$ is a function of the transmitted fractions, $\tau$, of the
$f_{scat}(\theta)$ and $f_{emit}$ fluxes measured with ($^\prime$) and without
the given screen:

\begin{equation}\label{marcuseq2}
\varphi = \frac{\tau^\prime_{scat}f_{scat}(\theta)+\tau^\prime_{emit}f_{emit}}{\tau_{scat}f_{scat}(\theta)+\tau_{emit}f_{emit}}
\end{equation}

\noindent We seek the diagnostic ratio $R_f(\theta) = f_{scat}(\theta)/f_{emit}$, which
relates to the comet grain albedo as $A(\theta) = R_f(\theta)/[1 +
R_f(\theta)]$.  Substituting in the equation above and solving,

\begin{equation}\label{marcuseq3}
R_f(\theta) =
\frac{\tau_{emit}\varphi - \tau^\prime_{emit}}{\tau^\prime_{scat} -
\tau_{scat}\varphi}.
\end{equation}

\noindent The four transmissivity coefficients are given in general form by,

\begin{equation}\label{marcuseq4}
\tau = \frac{\int\limits_{0}^{\infty} f(\lambda)\tau(\lambda)d\lambda}{\int\limits_{0}^{\infty} f(\lambda)d\lambda}
\end{equation}


\noindent where the generic $f(\lambda)$ is $f_{scat}(\theta)$ or $f_{emit}$ at a given
wavelength, $\lambda$, and
\begin{equation}\label{marcuseq5}
\tau(\lambda) = \tau_{atm}(\lambda)\tau_{opt}(\lambda)\tau_{scr}(\lambda)\tau_{rad}(\lambda)
\end{equation}

\noindent is the wavelength-specific transmissivity of the entire system, a product of the
individual wavelength-dependent transmissivities of the earth's atmosphere,
$\tau_{atm}(\lambda)$, the telescope optics, $\tau_{opt}(\lambda)$, the given
screen filter, $\tau_{scr}(\lambda)$, and the radiometer, $\tau_{rad}(\lambda)$.

To reduce the data, I constructed a comprehensive numerical model, details of
which will be given elsewhere.  Here it is sketched in broad outline.
$\tau_{atm}(\lambda)$ and $\tau_{opt}(\lambda)$ curves and models were obtained
from the standard literature or derived.  $\tau_{scr}(\lambda)$ curves were
obtained from \citet{1923Coblentz}
and correspondence in the Lowell Observatory archives.  $f_{scat}(\theta)$ and
$f_{emit}$ were modeled as Planckian black bodies with dust color and grain
superheat/silicate emission, respectively, as free parameters.  Free parameters
for $\tau_{atm}(\lambda)$ were water vapor, ozone and dust concentrations; for
$\tau_{opt}(\lambda)$, the effective thickness of tarnish on the silver mirror
coatings convolved with silver reflectivity; and for $\tau_{rad}(\lambda)$, the
efficiency (effective thickness) of the thermocouple blackener convolved with
the transmissivity of the radiometer rock salt window.  All transmissivities and
fluxes were incorporated in an Excel spreadsheet with a resolution of
$\lambda/\Delta\lambda = 200$. In the preliminary analysis presented here, I
restricted the computation of $R_f(\theta)$ to the water cell, pyrex, and quartz
screen data.  In a chi-square-related approach, I minimized summed squared
residuals in $R_f(\theta)$, as computed 
from $\varphi_{\rm H_2O}$, $\varphi_{\rm pyrex}$, and $\varphi_{\rm quartz}$, to
determine optimal values for the free parameters.

\section{Preliminary Results, and Further Work}

The fractional transmissions, $\varphi$, by the filter screens declined over the observing interval, 
most markedly for the water cell, as seen in the daily mean values in Table 1.  
The least-squares minimization disclosed that the thermocouple blackener efficiency was less 
than optimal, and that the grains had a high superheat.  Using optimized free parameters, 
the resulting mean daily $R_f(\theta)$ values are also provided in Table 1.  
Note the dramatic augmentation in $R_f(\theta)$ with decreasing $\theta$, with an over 10-fold (!) 
difference between Dec.\ 16 and 19.  This is due to {\it forward scattering} enhancement of 
the $f_{scat}(\theta)$ term in $R_f(\theta) = f_{scat}(\theta)/f_{emit}$ 
(see above) when the comet passed nearly between the Earth and Sun 
\citep{2007Marcus}.
That Lampland and the Sliphers could observe the comet in broad daylight owed substantially 
to this effect 
\citep{2007Marcus}.

Figure 2 plots the $R_f(\theta)$ values as large black diamonds, fitted to a phase function, 
$\Phi(\theta)$, which has been normalized to unity [$\log \Phi(\theta) = 0$] at $\theta = 90^\circ$.  
As comparisons, the large open symbols are fits for the two other comets, 
C/1975 V1 (West) and C/1980 Y1 (Bradfield), for which there is photometry of the scattered and 
thermal radiation in forward-scattering geometry (in these cases by broadband methods).  
In contrast, the small open symbols plot normalized photometry of solely $f_{scat}(\theta)$ 
from the LASCO C3 coronograph aboard the SOHO satellite, taken through a clear filter 
(0.4 $\mu$m -- 1.0 $\mu$m transmission) for comet 96P/Machholz, and a near-IR filter 
(0.8 $\mu$m -- 1.0 $\mu$m transmission) for C/2004 F4 (Bradfield).  
The SOHO data are less robust as proxies for $\Phi(\theta)$ than scattered/thermal 
photometry because of the need to assume the behavior of the baseline 
($\theta$-independent) brightness of the comets over the brief observing intervals.  
For further details and references, see 
\citet{2007Marcus}.
From Figure 2 we see that 1) Lampland's radiometric observations of comet C/1927 X1 
are concordant with the photometry of the other four comets, and 
2) all of the photometry fits the compound Henyey-Greenstein (HG) phase function model 
curves (for dust-to-gas light ratios of 1 and 10) well 
\citep{2007Marcus}.

\begin{figure}[t!]
\plotfiddle{'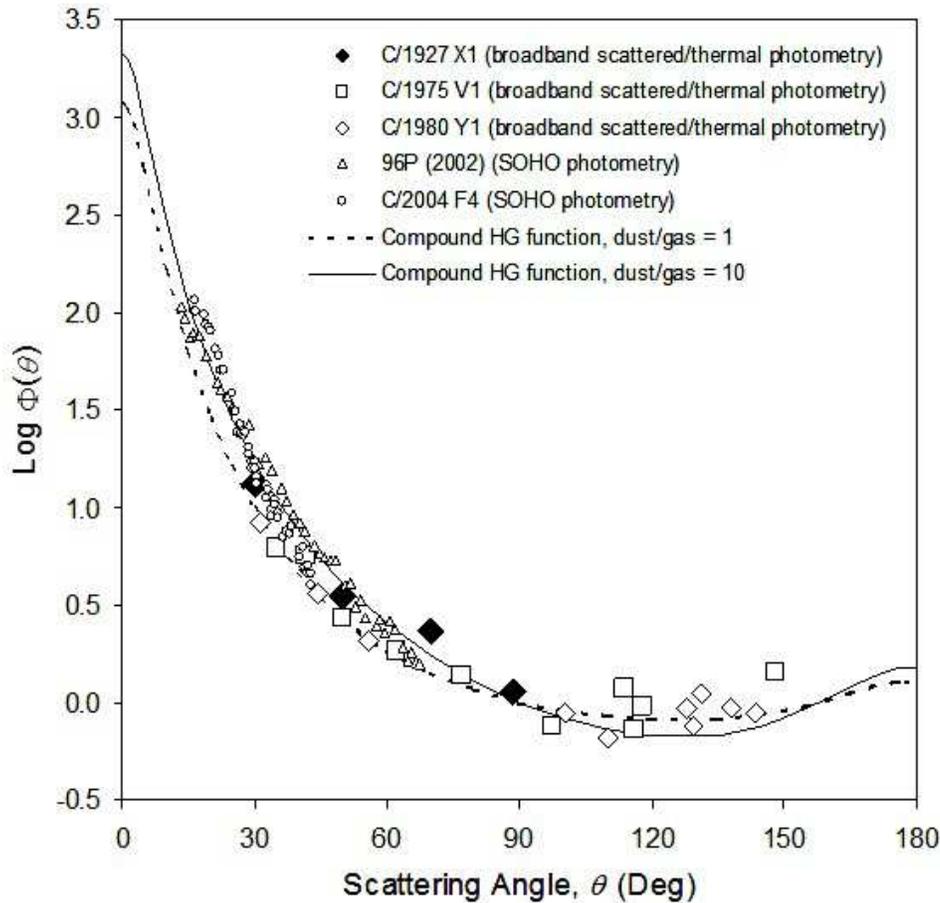'}{8cm}{0}{85}{85}{-175}{-100}
\vskip 1.3truein
\caption{
Forward-scattering enhancement of brightness as a function of scattering angle.  
Lampland's reduced radiometric observations of C/1927 X1 (large black diamonds) are 
concordant with the photometry for the four other comets, and the data for all five comets 
fit the model phase function curves well (see text).  Note that the scale for the y-axis is logarithmic.  
Modified from 
\citet{2007Marcus},
with permission of the {\it International Comet Quarterly} (Cambridge, Massachusetts).
}
\end{figure}

From this preliminary work, I conclude that the Lampland radiometry of comet C/1927 X1 is reducible and usable.  
The data reduction is ongoing, but I expect that further refinements to the comet's data points in 
Figure 2 will be small.  I have recently uncovered some unpublished transmission screen curves 
from the Lowell Observatory archives, which will be incorporated into the model so that the flux data 
for the full set of the filters can then be reduced.  Sodium emission by the comet is another issue.  
SOHO spacecraft photometry of sungrazing comets 
\citep{2002Biesecker,2010Knight}
indicates that sodium D line emission dominates the visual brightness of comets very close to the Sun, 
and my synthesis of the literature on quantitative cometary Na emission 
\citep{2010Marcus}
suggests 
that at the heliocentric distance of C/1927 X1 during Lampland's observations, Na brightness 
should have been comparable to that of the scattered sunlight.  Sodium brightness will 
therefore be incorporated as an additional model free parameter in the final reduction of the data.

\section{Parallels Between Lampland's Observations and V.~M. Slipher's Discovery of the Expanding Universe}

There are striking parallels between the first infrared observation of a comet by Lampland and 
Slipher's discovery of the recession of spiral nebulae.
\begin{itemize}
\item Each achievement was pioneering.  
\item Each was made possible by pioneering and state-of-the-art instrumentation.  
In Lampland's case this was the stellar radiometer, a sensitive but difficult instrument designed 
by his collaborator Coblentz and deployed at Lowell Observatory to measure the temperatures of 
the planets 
\citep{1923Coblentz}.
In Slipher's case, it was the Brashear spectrograph 
\citep{1904Slipher},
also a difficult and fussy instrument, but fast and capable of great resolution.
\item Each achievement was presented to American Astronomical Society (AAS) meetings to rave reviews.  
The abstract papers on C/1927 X1 
\citep{1928Lampland,1928Slipher}
were read on their behalves 
at the 1927 New Haven meeting by Roger Lowell Putnam (1893-1972), 
the newly-appointed sole trustee of Lowell Observatory.  John C.\ Duncan, Wellesley College, 
who was in attendance, wrote Lampland, ``These papers were received with enthusiasm and were the 
sensation of the meeting'' 
\citep{1928Duncan}.
Slipher read his paper on the mainly red-shifted velocities of the spiral nebulae at the 1914 
meeting in Evanston, Illinois.  
After he finished, he received an unprecedented standing ovation 
\citep{1970Hall}.
\item Each AAS presentation was published as an extended abstract in the organization's outlet at that time, 
{\it Popular Astronomy} 
\citep{1928Lampland,1915Slipher}.
\item Yet each achievement fell into relative obscurity with priority credit to fall to others.  
By the end of the 20th century, the infrared astronomy community was mistakenly touting C/1965 S1 
(Ikeya-Seki) as the first comet to be observed in the IR (see Introduction).  
And it is Edwin Hubble, who derived a coefficient (the ``Hubble'' constant) for the expansion 
of the Universe 
\citep{1929Hubble},
who so often gets the credit for the ``discovery'' of the expansion, 
which arguably should be shared with Lema\^itre and Slipher (see contributions by 
Peacock, Nussbaumer, Belenkiy, O'Raifeartaigh, and Way in this volume for more details).
\end{itemize}

\section{The Problematic Publication Culture at Lowell Observatory in the Early 20th Century}

To amplify on the final point, I suggest that a common thread for why each astronomer has received 
less credit than they deserve for their respective achievements was their attitude toward publication.  
In Lampland's case with C/1927 X1, the wound was clearly self-inflicted: he failed to follow up his 
sketchy abstract 
\citep{1928Lampland}
with a formal paper to exposit his pioneering data.  This behavior was by no means isolated for Lampland.  
His collaborator Coblentz, for example, nagged him to publish their further radiometry on the planets, 
and his failure to do so resulted in a major falling out between the two in 1926 
\citep{1980Hoyt}.

Why did Lampland not publish?  The common and sufficient explanation is that he was a perfectionist 
who was loath to stick his neck out 
\citep{1994Putnam,1987Smith,2007Tenn}.
``If 50 observations would do it, he would want 500,'' his Lowell colleague Henry Giclas (1910-2007) 
recounted recently 
\citep{2003Giclas}.
In addition, I believe that Lampland's hesitance with the C/1927 X1 data may have been cemented by the 
wild shift in the flux ratios over the four day observing period, reflected in the $\varphi_{H_2O}$ values in Table 1.
Such behavior would have defied explanation -- he and Coblentz had never encountered anything of that 
sort in their radiometry of the planets 
\citep{1923Coblentz}.
Astronomers of the day appeared to be unaware of the forward-scattering property of comet grains 
\citep{2007Marcus},
which greatly enhances the scattered (visible and near-IR) flux at small scattering angles.  
This phenomenon would have offered Lampland a framework for explaining the flux ratio changes in his comet data.

In Slipher's case, his seminal redshift data on the spiral nebulae were published in {\it Popular Astronomy} 
\citep{1915Slipher}
and the rather peripheral {\it Proceedings of the American Philosophical Society} 
\citep{1917Slipher},
rather than in the leading mainline astronomical journals of the day like {\it Astrophysical Journal}, 
{\it Astronomical Journal}, or {\it Astronomische Nachrichten}.  An audit of the Lowell Observatory 
archives and Astrophysical Data Service (ADS) documents that the publication records of Slipher and 
Lampland, especially after Percival Lowell's death in 1916, were fairly sparse.  Much of what both 
Slipher and Lampland did publish was confined internally to the {\it Lowell Observatory Bulletin}, 
an arguably insular and suboptimal route for obtaining wider recognition for their work, although 
admittedly observatory bulletins were a common outlet for publication in those days.

Slipher's and Lampland's approaches to publication were not simply lackadaisical; in some respects they 
were even negative.  The young and ambitious Arthur Adel (1908-1994), at Lowell Observatory in the 1930s, 
relates having to keep his data from Lampland for fear that they would be sequestered and never published 
\citep{1987Smith}.
Adel also relates having to submit his manuscripts to Slipher for clearance, and then battling him to get 
them sent off to the journals 
\citep{1987Smith}.

If Slipher was injured by Hubble's failure to cite his spiral nebula redshift work in the seminal paper 
on the linear dependence of the recession velocities on distance 
\citep{1929Hubble}
-- a theme of this conference -- then it is also apparent that Slipher and Lampland did themselves 
no favors by their approaches to publication.  The Lampland comet incident, as presented here, 
illuminates this point.

\acknowledgements I thank Donald K.\ Yeomans, Jet Propulsion Laboratory, for providing his correspondence 
with the late Arthur Hoag, Antoinette Beiser, Lowell Observatory (LO), for aid in archives research, 
Henry Giclas (1910-2007), LO, for information on circumstances of the observations, 
Giclas and Kevin Schindler, LO, for details about the telescope, and two
anonymous reviewers and the Editors for their very helpful suggestions.

\bibliography{marcus}

\end{document}